\documentclass[mathleft
]{an}
\usepackage{graphicx}
\usepackage{times}
\usepackage{natbib}
\bibpunct{(}{)}{;}{a}{}{,}
\newcommand{\ro}{{\it ROSAT}}
\newcommand{\xmm}{{\it XMM-Newton}}
\newcommand{\euclid}{{\it Euclid}}
\newcommand{\rosi}{{\it eROSITA}}
\newcommand{\srg}{{\it Spektr-RG}}
\newcommand{\athena}{{\it Athena}}
\newcommand{\cha}{{\it Chandra}}
\newcommand{\eehifl}{{\emph{eeHIFLUGCS}}} %
\newcommand{\msun}{M_{\odot}}
\begin{document}

% The following seven commands are intended for editorial usage and should be ignored by
% the author(s).
\Pagespan{789}{}% Document's page range. 
% If second parameter is left empty, the last page is computed automatically.
\Yearpublication{2017}%
\Yearsubmission{2016}%
\Month{NN}%   
\Volume{NNN}%  
\Issue{NN}% 
% \DOI{This.is/not.aDOI}% 

\title{Follow-up of \textit{eROSITA} and \textit{Euclid} Galaxy Clusters with \textit{XMM-Newton}}

\author{T.H. Reiprich\inst{1}\fnmsep\thanks{Corresponding author:
  \email{reiprich@astro.uni-bonn.de}\newline}
}
\titlerunning{Instructions for authors}
\authorrunning{T.H. Reiprich}
\institute{
Argelander Institute for Astronomy, Bonn University,
Auf dem H\"ugel 71, 53121 Bonn, Germany
}

\received{01 Oct 2016}
\accepted{30 Nov 2016}
\publonline{later}


\abstract{%
A revolution in galaxy cluster science is only a few years away. The survey machines \rosi\ and \euclid\ will provide cluster samples of never-before-seen statistical quality. \xmm\ will be the key instrument to exploit these rich datasets in terms of detailed follow-up of the cluster hot gas content, systematically characterizing sub-samples as well as exotic new objects.
  }

\maketitle

\section{Introduction}
Do we still need \xmm\ in $\sim$2020--2030 for galaxy cluster and cosmology studies? The answer is a very clear ``yes!'' because (i) \rosi\ and \euclid\ will discover tens to hundreds of thousands of new clusters, revealing rare, exotic ones with features never before seen and worthy of detailed study; (ii) systematic follow-up of subsamples is required to characterize and understand these large samples, for both cluster astrophysics and cosmology; (iii) \xmm\ will stay the most efficient instrument for hot ($\sim$5--10 keV) clusters; (iv) these new clusters will start flowing in end of 2018 -- and they will keep flowing for a decade.

\section{What will \rosi\ and \euclid\ provide?}
\subsection{\rosi}
\rosi\ is the primary instrument to be launched aboard \srg\ end of 2017 \citep[e.g.,][]{pab14}. It will perform an X-ray all-sky survey more than one order of magnitude more sensitive than the \ro\ All-Sky Survey. The main science driver is to constrain the properties of dark energy with galaxy clusters but breakthroughs can be expected for almost all other astrophysical objects as well \citep[e.g.,][]{mpb12}. About 100,000 clusters are expected to be discovered and used for cosmological tests, with a median redshift $z\sim0.3$ and a tail extending to $z>1$ \citep[e.g.,][]{ppr12}. Typical cluster masses at the median redshift will be a few times $10^{14}\msun$ \citep[e.g.,][]{brm14}. A constant dark energy equation of state is expected to be constrained to $\sim$3\% and the 2$\sigma$ figure-of-merit (FoM) introduced in the Dark Energy Task Force (DETF) report \citep{abc06} reaches about 55 when combined with Planck constraints (Pillepich et al., in prep.; Borm et al., in prep.).
These forecasts assume cluster redshifts to be available and the X-ray luminosity--mass scaling relation from \citet{vbe09} with uncertainties reduced by a factor of four, as optimistically assumed to be available at the end of the \rosi\ survey ($\sim$2022).

\subsection{\euclid}
\euclid\ is a medium class mission (M2) in ESA's Cosmic Vision program expected to be launched end of 2020. Its primary science driver is the study of dark energy mainly using gravitational lensing (cosmic shear) and baryon acoustic oscillations (BAOs) by performing a visible/near-infrared imaging-spectroscopy survey over $\sim$15,000 sq.deg \citep[e.g.,][]{laa11}. It is designed to achieve a (1$\sigma$; i.e., different from the above DETF definition) FoM $>$400.
Additional science includes the detection of up to $10^6$ galaxy clusters out to $z\sim2$ and their use for cosmology \citep[e.g., assuming perfectly known scaling relations, 1$\sigma$ FoM $>$800,][]{sbf16}.

\section{How will \xmm\ be useful?}
\begin{figure*}
\includegraphics[width=160mm]{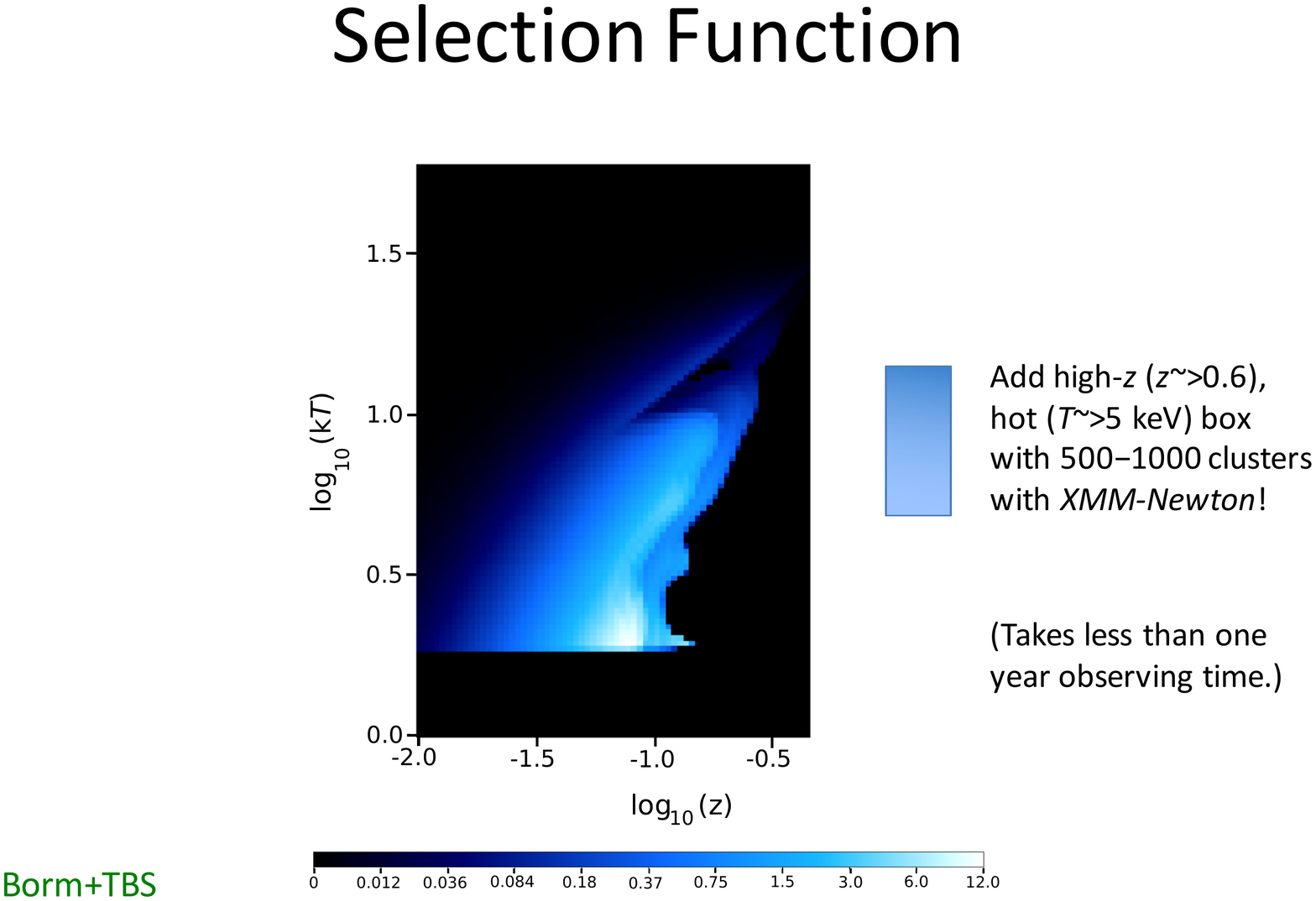}
\caption{Preliminary expected selection function of \rosi\ clusters with measured temperatures (\citealt{brm14}; Borm et al., in prep.). The x-axis shows the logarithmic redshift and the y-axis the logarithmic measured (from simulations) X-ray temperature in units of keV. The color scale shows the number of clusters per pixel; i.e., black means zero clusters and white about 12 clusters per bin. The blue box to the right qualitatively illustrates the selection region for dedicated \xmm\ follow-up of 500--1,000 clusters expected to improve cosmological constraints significantly. Neither the location nor the color of this box is meant to be interpreted quantitatively.}
\label{T-select}
\end{figure*}

An important ingredient for the \rosi\ cosmology analysis is the X-ray luminosity--mass scaling relation, which exhibits significant intrinsic scatter, and this applies similarly to galaxy-based mass estimators to be employed for the \euclid\ analysis. A large scatter generally implies that measurement uncertainties for the parameters used to model the relation and its evolution based on small cluster samples are also large. The scatter of the X-ray temperature--mass relation, on the other hand, is comparatively small. Potentially, even tighter cosmological constraints could be obtained if gas temperatures were available (e.g., Borm et al., in prep.). However, in this case, the limiting factor for cosmological power quickly becomes the reduction in cluster numbers since temperature measurements are about an order of magnitude more costly in terms of required exposure time than X-ray luminosity; for example, the \rosi\ survey itself will provide temperatures for ``only'' about 2,000 clusters, and only at low redshift (up to $z\sim0.15$) where the leverage on the evolution of the cluster number density is small \citep[Fig.~\ref{T-select}, and ][]{brm14}. At low redshift, the cluster number density does not depend strongly on the dark energy equation of state, so improvements on those constraints from \rosi-only temperature measurements will be small.

\begin{figure}
\includegraphics[width=80mm]{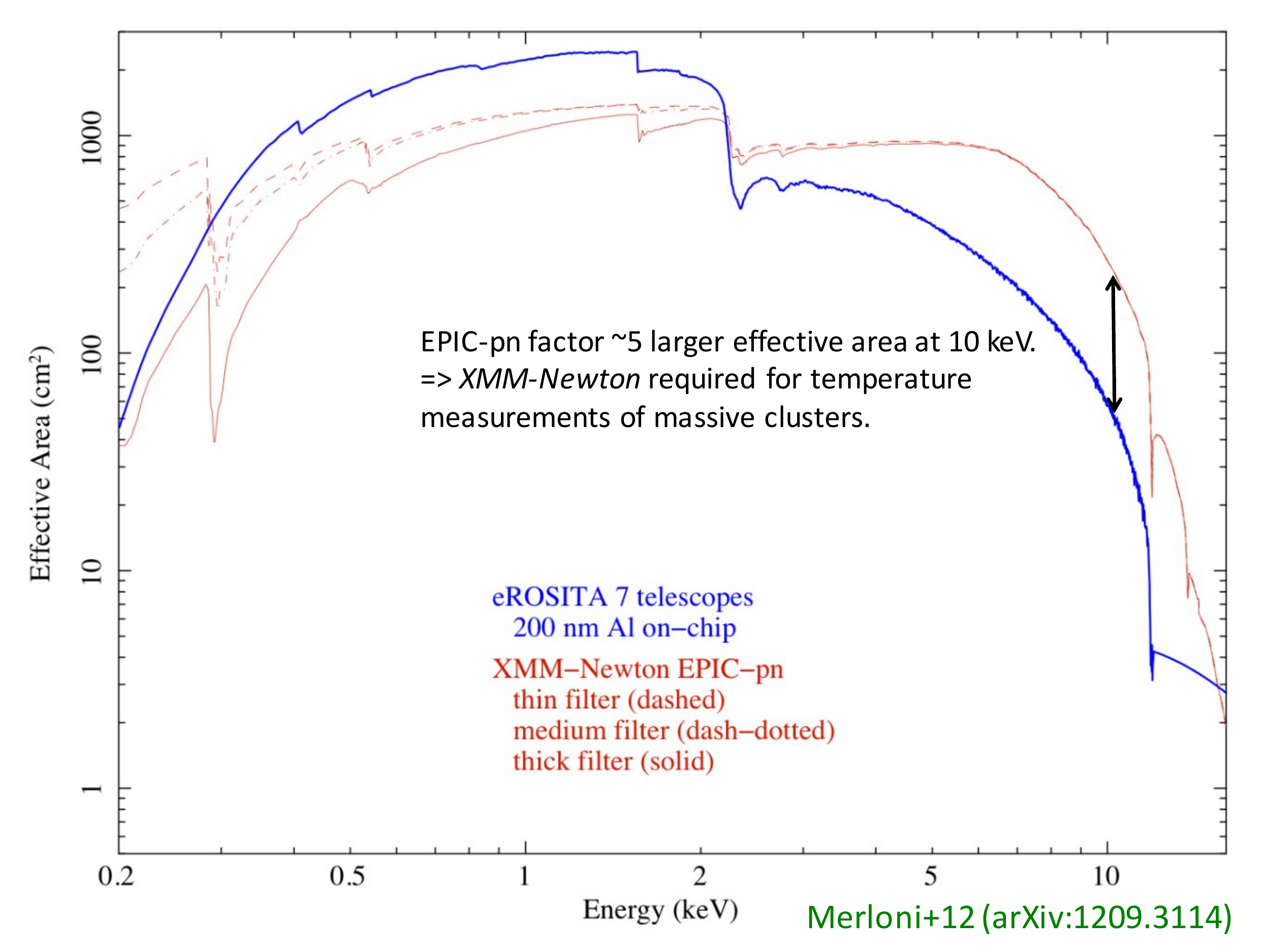}
\caption{Comparison of \xmm\ EPIC-pn and \rosi\ effective area as a function of photon energy. While \rosi\ outperforms EPIC-pn at soft energies, \xmm\ is significantly more efficient at high energies ($>$5 keV). Adapted from \citet{mpb12}.}
\label{eff_area}
\end{figure}

This is where \xmm\ comes into play. It could be used to provide gas temperatures of hot ($>$5 keV) clusters at higher redshifts ($>$0.6) (approximated by the blue box in Fig.~\ref{T-select}) missing from the direct \rosi\ measurements and allowing us to constrain the cluster redshift evolution. \xmm\ will likely stay the most efficient instrument to measure gas temperatures of hot clusters until the launch of ESA's L2 large class mission \athena\ in 2028 due to its large effective area at high photon energies (Fig.~\ref{eff_area}). Of course, combining temperatures measured with \rosi\ and \xmm\ requires a significant cross-calibration effort to quantify systematic differences accurately; but this can be done as has already been demonstrated for the \cha\ and \xmm\ instruments \citep[e.g.,][]{srl14}. 

\begin{figure*}
\includegraphics[width=160mm]{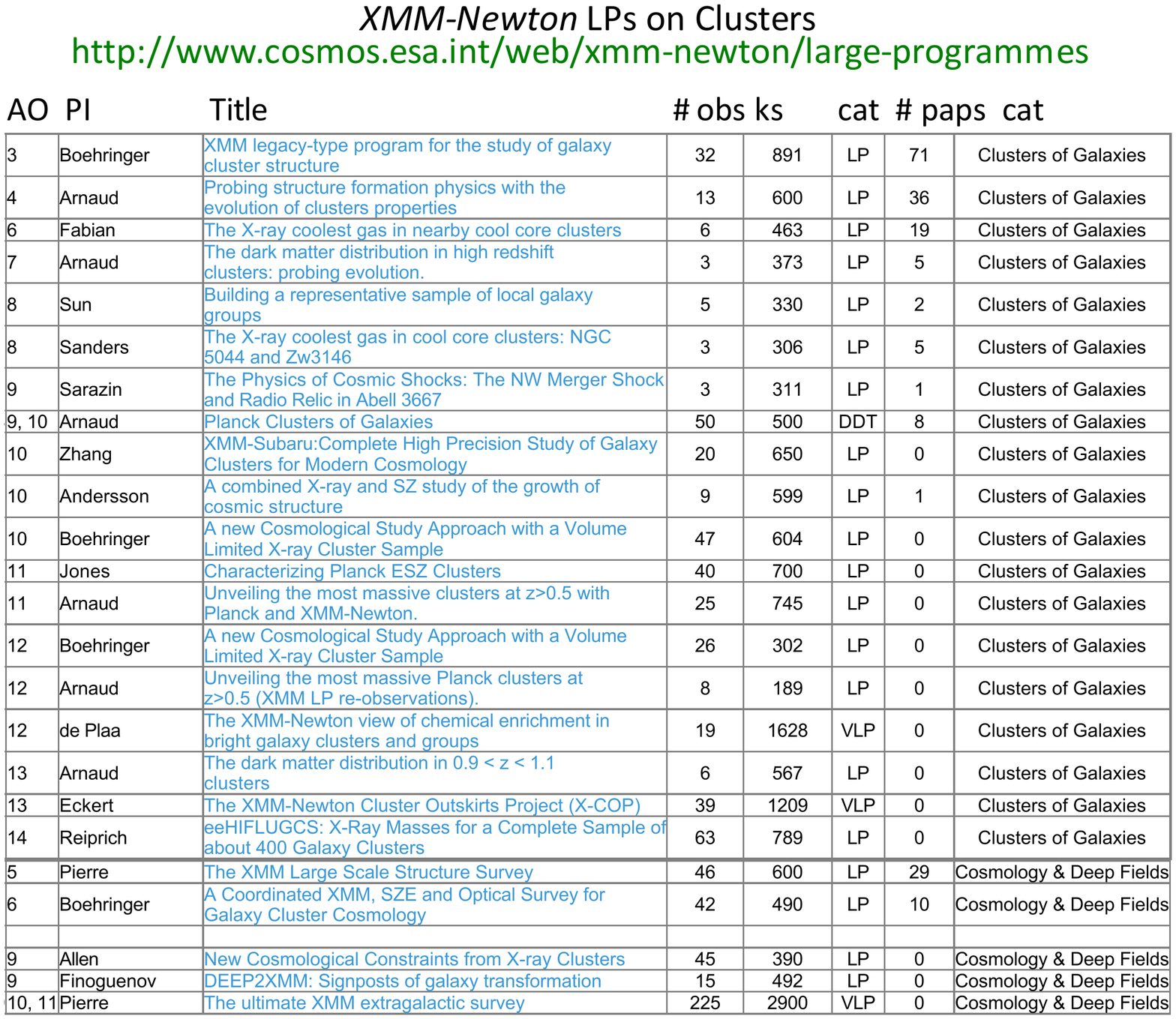}
\caption{LPs and VLPs. Based on http://www.cosmos.esa.int/web/xmm-newton/large-programmes as of May 2016.}
\label{VLPs}
\end{figure*}
\begin{figure*}
\includegraphics[width=160mm]{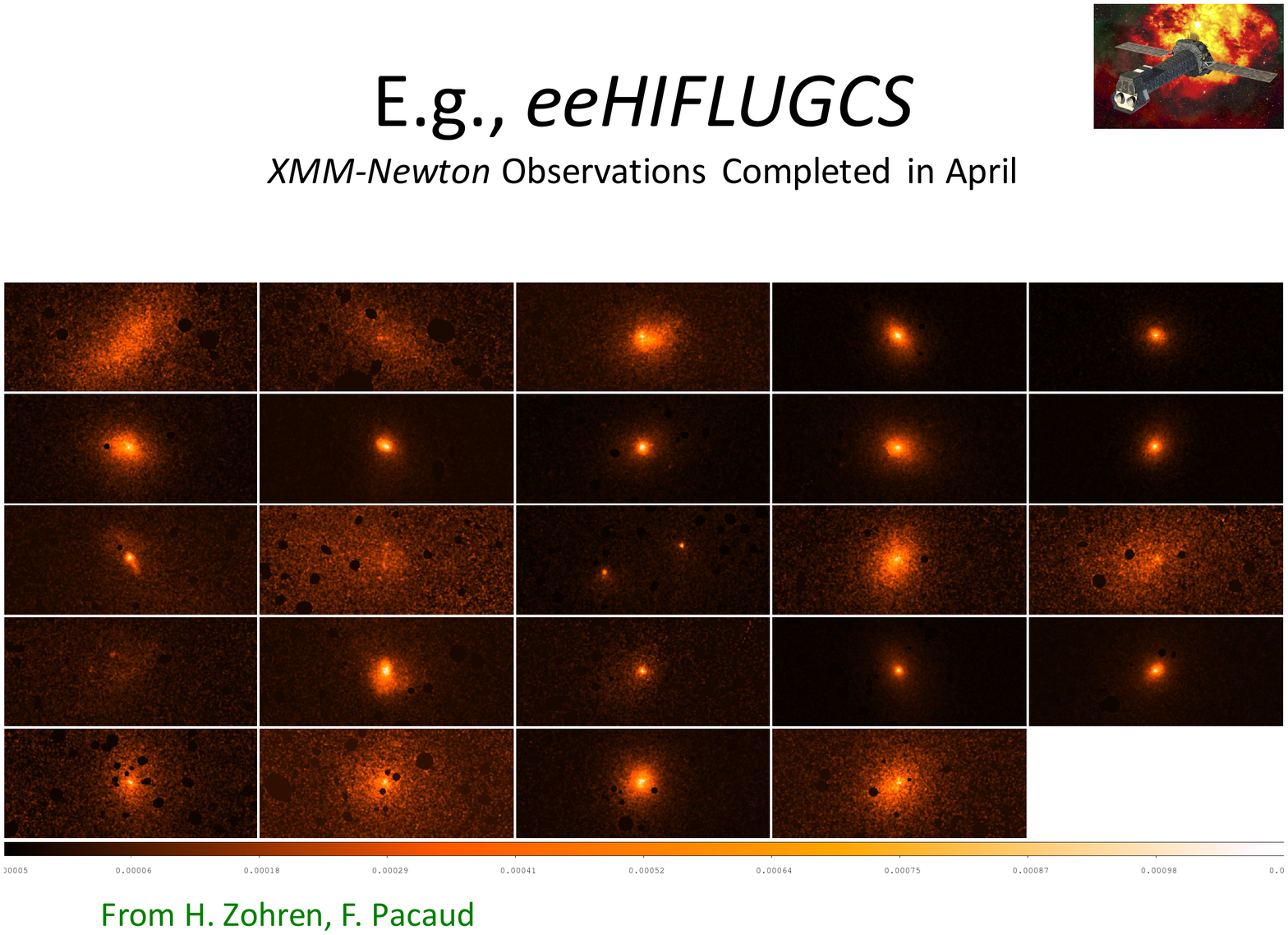}
\caption{Some \eehifl\ clusters recently observed with \xmm. Notice the great variety of morphologies -- e.g., sloshing and merging signatures, double clusters, some centrally peaked but also a number of low surface brightness objects -- that appears when following-up complete samples down to low fluxes. Reduction and analysis by H. Zohren and F. Pacaud. The \xmm\ image on the upper right comes from the \xmm\ website.}
\label{eehifl}
\end{figure*}
Would such an \xmm\ program be feasible in practice? The first (out of 8 in total) all-sky survey will be completed $\sim$9 months after the \rosi/\srg\ launch. A robust cluster catalog including redshifts is currently expected to be available about two years later. Therefore, for a launch end of 2017, a first multi-cycle \xmm\ program could be proposed around October 2020. A proposal extending to fainter clusters could come a few years later, if the initial results look promising. While quantitative projections are not yet available, a total number of 500--1,000 hot clusters with precise temperature measurements beyond $z\sim0.6$ is expected to make a big qualitative difference in terms of cosmological constraints. Many more such clusters will actually be available from the \rosi\ catalog (e.g., more than 1,000 hot clusters are expected even beyond $z\sim1$, \citealt{ppr12}, Section 9.3). A thorough feasibility study will have to be carried out for such a proposal at the time, of course; for an educated guess, one can consider that precise single temperature measurements of massive higher redshift clusters require about 10--20 ks. Performing such measurements for 500--1,000 clusters, therefore, takes about 10 Ms; i.e., less than one year of typical \xmm\ time per Announcement of Opportunity (AO; $\sim$14 Ms). More realistically, adding a soft-proton-flare-factor ($\sim$1.4), this comes out to pretty much one year in total. However, this would naturally be spread over multiple cycles; e.g., for five cycles, this would amount to $\sim$20\% of the total time. So, per AO, this would even fit well within the existing Very Large Programme category (up to 3 Ms).

In addition to the specific cosmology example above, \xmm\ will play a key role for other galaxy cluster science areas. To provide some background for this, let us have a look at previous Large and Very Large Programmes (LPs, VLPs) performed with \xmm\ in the field of galaxy clusters. Fig.~\ref{VLPs} lists LPs and VLPs with a relation to galaxy clusters according to the \xmm\ website.
As is obvious from the titles and number of observations, the large majority of (V)LPs observe \textit{samples} of clusters, typically selected in a well-defined way from larger parent samples. Such follow-up observations of complete sub-samples (e.g., \eehifl, Fig.~\ref{eehifl}) often serve two purposes: (i) to constrain population properties in a systematic way with corrections for selection effects, and (ii) to discover rare, new, unexpected, and exciting phenomena.

An example for the latter has been provided by \citet[][see, e.g., http://chandra.harvard.edu/photo/2015/z8338/]{sr15}. In systematic sample follow-up observations, they discovered a large galaxy, apparently falling in or passing by a cluster, exhibiting one of the longest X-ray tails ever observed for a single galaxy. Even more surprisingly, the tail has been \textit{completely} stripped off the galaxy, including the high density head. Further theoretical and observational studies of this system will provide detailed insights into stripping processes and gas microphysics.

Naively, one might think one has to be quite lucky to run into such rare systems. That is, however, not the case. In order to systematically discover phenomena that occur, say, in only $\sim$1\% of the population, one simply has to follow-up in detail $>$100 objects. With \xmm\ and \cha\ we have now entered the era where 100s of well-selected objects are indeed being followed up. In the future, given the new large parent samples provided by \rosi\ and \euclid, we will be able to find striking objects that occur only once in $>$1,000 clusters!

\section{Suggestions for future \xmm\ observing strategies}
Based on the above assessment and examples, some suggestions for modifications (or not) of the current \xmm\ observing strategies are outlined below.
\begin{itemize}
 \item Improve the possibility for very large systematic studies of statistically meaningful, representative, and complete samples further; e.g., $>$3 Ms programs, maybe split over $>$2 cycles.
 \item Keep the possibility to study individual interesting targets with short ($\ll$100 ks) and long exposures ($\gg$100 ks). Such targets may, e.g., be found through systematic follow-up observations of large samples.
 \item Encourage Time Allocation Committees (TACs) to look more favorably upon completing ``the last 10\% of the sample.'' The naive $\sqrt{N}$ argument often does not apply to well-selected samples. For instance, the unobserved 10\% will likely not be drawn from the same parent population as the full sample if the follow-up plan includes a large fraction of independent archival observations (i.e., observations of ``interesting'' individual objects proposed because of their special properties). Therefore, an unknown selection bias would be introduced that cannot be corrected for.
 \item Require a minimum (flare-corrected) exposure of $>$6 ks, unless specifically justified, e.g., for variability studies, since \rosi\ will provide $\sim$1.6 ks anywhere.
 \item Increase the efficiency by observing low-surface brightness objects at low-flare-risk periods. For example, typically, the exposure for all flared times is lost for low-surface brightness objects, often resulting in science being partially or completely lost, while this is not necessarily the case for very bright point sources.
\end{itemize}

\acknowledgements
THR acknowledges support by the German Research Association (DFG) through grant RE 1462/6 and through the Transregional Collaborative Research Centre TRR33 The Dark Universe (project B18) as well as by the German Aerospace Agency (DLR) with funds from the Ministry of Economy and Technology (BMWi) through grant 50 OR 1608.

\newpage%

\end{document}